\documentclass[twocolumn,prb,showpacs,preprintnumbers,amsmath,amssymb]{revtex4}

\usepackage{graphicx}
\usepackage{dcolumn}
\usepackage{bm}
\usepackage{color}
\begin{document}


\title{Pauli-limited superconductivity and antiferromagnetism in the heavy-fermion compound CeCo(In$_{\bm{1-x}}$Zn$_{\bm x}$)$_{\bm{5}}$}

\author{Makoto~Yokoyama}
\email[Electronic Address: ]{makoto.yokoyama.sci@vc.ibaraki.ac.jp}
\author{Hiroaki~Mashiko}
\author{Ryo~Otaka}
\author{Yumi~Sakon}
\author{Kenji~Fujimura}
\affiliation{Faculty of Science, Ibaraki University, Mito 310-8512, Japan}
\author{Kenichi~Tenya}
\affiliation{Faculty of Education, Shinshu University, Nagano 380-8544, Japan}
\author{Akihiro~Kondo}
\author{Koichi~Kindo}
\author{Yoichi~Ikeda}
\author{Hideki~Yoshizawa}
\author{Yusei~Shimizu}
\author{Yohei~Kono}
\author{Toshiro~Sakakibara}
\affiliation{Institute for Solid State Physics, The University of Tokyo, Kashiwa 277-8581, Japan}
\date{\today}
             
\begin{abstract}
We report on the anisotropic properties of Pauli-limited superconductivity (SC) and antiferromagnetism (AFM) in the solid solutions CeCo(In$_{1-x}$Zn$_x$)$_5$ ($x\le0.07$).  In CeCo(In$_{1-x}$Zn$_x$)$_5$, the SC transition temperature $T_c$ is continuously reduced from 2.3 K ($x=0$) to $\sim 1.4\ {\rm K}$ ($x=0.07$) by doping Zn, and then the AFM order with the transition temperature of $T_N \sim 2.2\ {\rm K}$ develops for $x$ larger than $\sim 0.05$. The present thermal, transport and magnetic measurements under magnetic field $B$ reveal that the substitution of Zn for In yields little change of low-temperature upper critical field $\mu_0H_{c2}$ for both the tetragonal $a$ and $c$ axes, while it monotonically reduces the SC transition temperature $T_c$. In particular, the magnitudes of $\mu_0H_{c2}$ at the nominal Zn concentration of $x = 0.05$ (measured Zn amount of $\sim 0.019$) are 11.8 T for $B\,||\,a$ and 4.8 T for $B\,||\,c$, which are as large as those of pure compound though $T_c$ is reduced to 80\% of that for $x=0$. We consider that this feature originates from a combination of both an enhanced AFM correlation and a reduced SC condensation energy in these alloys. It is also clarified that the AFM order differently responds to the magnetic field, depending on the field directions. For $B\,||\,c$, the clear anomaly due to the AFM transition is observed up to the AFM critical field of $\sim 5\ {\rm T}$ in the thermodynamic quantities, whereas it is rapidly damped with increasing $B$ for $B\,||\,a$. We discuss this anisotropic response on the basis of a rich variety of the AFM modulations involved in the Ce115 compounds.
\end{abstract}

\maketitle

\section{Introduction}
Quantum critical phenomena are presently one of the central issues in heavy-fermion physics. They emerge in the proximity of the phase transition at zero temperature called the quantum critical point (QCP), which is generated by suppressing magnetically ordered phases by applying pressure, a magnetic field, and chemical substitution. In particular, unconventional superconductivity (SC) often occurs in the vicinity of the antiferromagnetic (AFM) QCP,\cite{rf:Mathur98,rf:Pfleiderer2009} strongly suggesting that the AFM fluctuation enhanced near the QCP is crucial for Cooper pairing between the correlated electrons in the SC order. It is also expected that the quantum fluctuation enhanced near the QCP yields a new class of low-energy excitations of the conduction electrons different from the Fermi-liquid quasiparticle state. In fact, most heavy-fermion compounds involving the quantum critical instability show the non-Fermi-liquid (NFL) behavior in thermal, magnetic and transport quantities at very low temperatures.\cite{rf:Stewart2001,rf:Lohneysen2007} 

The heavy-fermion superconductor CeCoIn$_5$ (the HoCoGa$_5$-type tetragonal structure) is one of the most intensively investigated heavy-fermion compounds from the perspective of the interplay between the quantum critical fluctuation and the SC state, as well as its anomalous SC properties.\cite{rf:Petrovic2001} It shows a SC transition characterized by the anomalously large specific-heat jump $\Delta C_p/\gamma T_c=4.5$ at $T_c=2.3\ {\rm K}$. The magnetically mediated pairing mechanism is inferred from the $d$-wave ($d_{x^2-y^2}$) symmetry of the SC gap evidenced by thermal-conductivity\cite{rf:Izawa2001} and specific-heat\cite{rf:An2010} measurements under rotated weak magnetic fields, and conductance measurement for junctions.\cite{rf:Park2008} A strong Pauli-paramagnetic effect leads to a first-order transition at the SC upper critical field $H_{c2}$ below 0.7 K,\cite{rf:Izawa2001,rf:Tayama2002,rf:Ikeda2001,rf:Bianchi2002} and another SC phase coexistent with an incommensurate AFM order (the so-called $Q$ phase) appears at very low temperatures below 0.3 K and high fields just below $H_{c2}$.\cite{rf:Bianchi2003-1,rf:Radovan2003,rf:Kakuyanagi2005,rf:Young2007,rf:Kenzelmann2008} Moreover, the NFL features are recognized as a weak divergence in magnetization,\cite{rf:Tayama2002} $-\ln T$ behavior in specific heat divided by temperature, and a $T$-linear dependence in electrical resistivity above $H_{c2}$.\cite{rf:Bianchi2003-2,rf:Paglione2003} These NFL features are interpreted as being the AFM quantum critical fluctuations evolving near the QCP. It is therefore suggested that the low-temperature physical properties of CeCoIn$_5$ including the SC order are inherently related to the AFM quantum criticality. Furthermore, this quantum critical behavior is considered to involve the multiple characteristics of the spin fluctuations.\cite{rf:Paglione2003}

The long-range AFM order is actually induced by the ionic substitution for CeCoIn$_5$. It is revealed in the mixed compounds Ce(Co,Rh)In$_5$ that replacing 20\% of Co by Rh generates the AFM order, accompanying a reduction of the SC transition temperature $T_c$ down to $\sim 1.5\ {\rm K}$.\cite{rf:Zapf2001,rf:Jeffries2005} The AFM-QCP is likely located near its critical Co concentration because the AFM transition temperature continuously increases with further doping Rh. The substitution of very small amount of Cd or Hg for In also yields a similar effect; doping Cd or Hg suppresses the SC phase, and then the AFM order develops above a few percent of the substitutions.\cite{rf:Pham2006,rf:Urbano2007,rf:Tokiwa2010,rf:Capan2010,rf:Gofryk2012} The AFM structure in these alloys involves the $q=(1/2,1/2,1/2)$ modulation,\cite{rf:Yoko2006,rf:Ohira-Kawamura2007,rf:Yoko2008,rf:Nicklas2007} and the coupling between the SC and the AFM is suggested by the observation of the spin-resonance excitation at the same $q$ vector in the SC phase of pure CeCoIn$_5$.\cite{rf:Stock2008,rf:Raymond2012} On the other hand, the propagation vector of the AFM order induced in the (Ce,Nd)CoIn$_5$ alloys is found to be incommensurate, corresponding to that seen in the $Q$ phase of pure CeCoIn$_5$.\cite{rf:Raymond2014} Quite recently, a precise inelastic neutron scattering experiment for pure CeCoIn$_5$ indicates that the inelastic spin-resonance mode has this incommensurate propagation vector rather than the commensurate one.\cite{rf:Raymond2015} CeCoIn$_5$ is thus suggested to possess a rich variety of coupling properties between the SC and the AFM, but their entire aspects have not yet been fully uncovered. 

To obtain the clues for understanding those aspects, low-temperature properties of new mixed compounds CeCo(In$_{1-x}$Zn$_x$)$_5$ have been investigated by performing macroscopic measurements. We recently made a brief report on the finding that doping Zn generates the AFM order below $T_N\sim 2.2$ K for $x$ larger than $\sim 0.05$ along with the reduction of $T_c$.\cite{rf:Yoko2014} In the previous report, however, the investigation using the magnetic field was limited in the isothermal magnetization measurement for the field applied along the tetragonal $c$ direction. Since in CeCoIn$_5$, the magnetic anisotropy is considered to strongly influence the Pauli-limited $H_{c2}$, it is interesting to investigate the anisotropic response of the SC order to the magnetic field, and its relationship to the AFM order in CeCo(In$_{1-x}$Zn$_x$)$_5$. In this paper, we describe the properties of the SC and AFM orders under both the $a$- and $c$-axis magnetic fields, which are investigated by means of magnetization, electrical resistivity and specific heat measurements. In particular, the present investigations reveal the magnetic-field versus temperature phase diagrams of the SC and AFM phases for the $a$- and $c$-axis field directions, in which the strong anisotropy of the AFM phase boundary and the evolution of the NFL behavior in the vicinity of its border are realized. Furthermore, the anisotropy of $H_{c2}$ is explored for the Zn concentrations of $x \le 0.07$. These experimental results are discussed on the basis of the strong Pauli paramagnetic effect and a rich variety of the AFM structures possibly involved in CeCo(In$_{1-x}$Zn$_x$)$_5$.

\section{Experiment Details}
Single crystals of CeCo(In$_{1-x}$Zn$_x$)$_5$ with $x=0$, 0.025, 0.05 and 0.07 were grown by means of the indium-flux technique. High-purity materials with a molar ratio of ${\rm Ce:Co:In:Zn}=1:1:(1-x)F:xF$ with $F=15-20$ were set and sealed in quartz tubes under 0.02 MPa Ar atmosphere. Note that we here choose the initial Zn concentrations as the ratio of the flux amounts in contrast to the stoichiometric amounts of Sn being prepared in the growth of CeCoIn$_{5-x}$Sn$_x$.\cite{rf:Bauer2005} They were heated up to 1050$^\circ$C and then cooled with the two-stage process similar to the previously reported one \cite{rf:Petrovic2001}; initially fast-cooled down to 800$^\circ$C and then slowly lowered to 500$^\circ$C. Plate-shaped single crystals with the basal plane perpendicular to the tetragonal $c$ axis were obtained after removing the excess flux using hydrochloric acid, and their tetragonal structure was checked by means of the x-ray diffraction technique. The energy dispersive x-ray spectroscopy (EDS) measurements for these samples indicate the homogeneous distributions of all the elements. However, the estimated Zn amounts were roughly 35\% ($\pm20\%$) of nominal $x$ values, and it is hard to determine the exact Zn concentrations using the usual EDS technique for the insufficient resolution in the EDS measurement compared with the smallness of the Zn amounts.\cite{rf:Yoko2014} It is known that such a deviation frequently happens in the samples grown by the flux technique, and indeed, the large deviations are also found in the isostructural CeCo(In,Cd)$_5$, CeCo(In,Hg)$_5$, and CeRh(In,Sn)$_5$ alloys.\cite{rf:Pham2006,rf:Bauer2006} We here use the nominal $x$ values throughout this paper for clarity and simplicity, and the actual Zn concentrations $y$ estimated by the EDS measurements are also shown along with the nominal values, except for the $y$ value for $x=0.025$ involving a fairly large uncertainty due to the extremely small Zn concentration.

Despite the large difference between nominal and actual Zn concentrations, there is little sample dependence on the low-temperature ordered phases for each $x$. In Fig.\ 1(a), we compare the low-temperature part of the electrical resistivity $\rho$ for $x=0.07$ ($y\sim 0.025$), obtained from the various samples grown separately. These data show a clear kink at $T_N=2.2\ {\rm K}$ and a drop at $T_c = 1.4\ {\rm K}$ due to the AFM and SC transitions, respectively, and we find that they reproduce fairly well except for slight differences of about 0.1 K in $T_c$ and $T_N$. As displayed in Fig. 1(b), a cusp seen in temperature variations of magnetization is ascribed to the AFM transition, and the SC order brings about a large drop of ac susceptibility due to the shielding effect. In addition, specific heat shows clear jumps associated with these transitions [Fig.\ 1(c)]. The above features are consistent with the previously reported ones.\cite{rf:Yoko2014} 
\begin{figure}[tbp]
\begin{center}
\includegraphics[bb=7 451 324 824,keepaspectratio,width=0.5\textwidth]{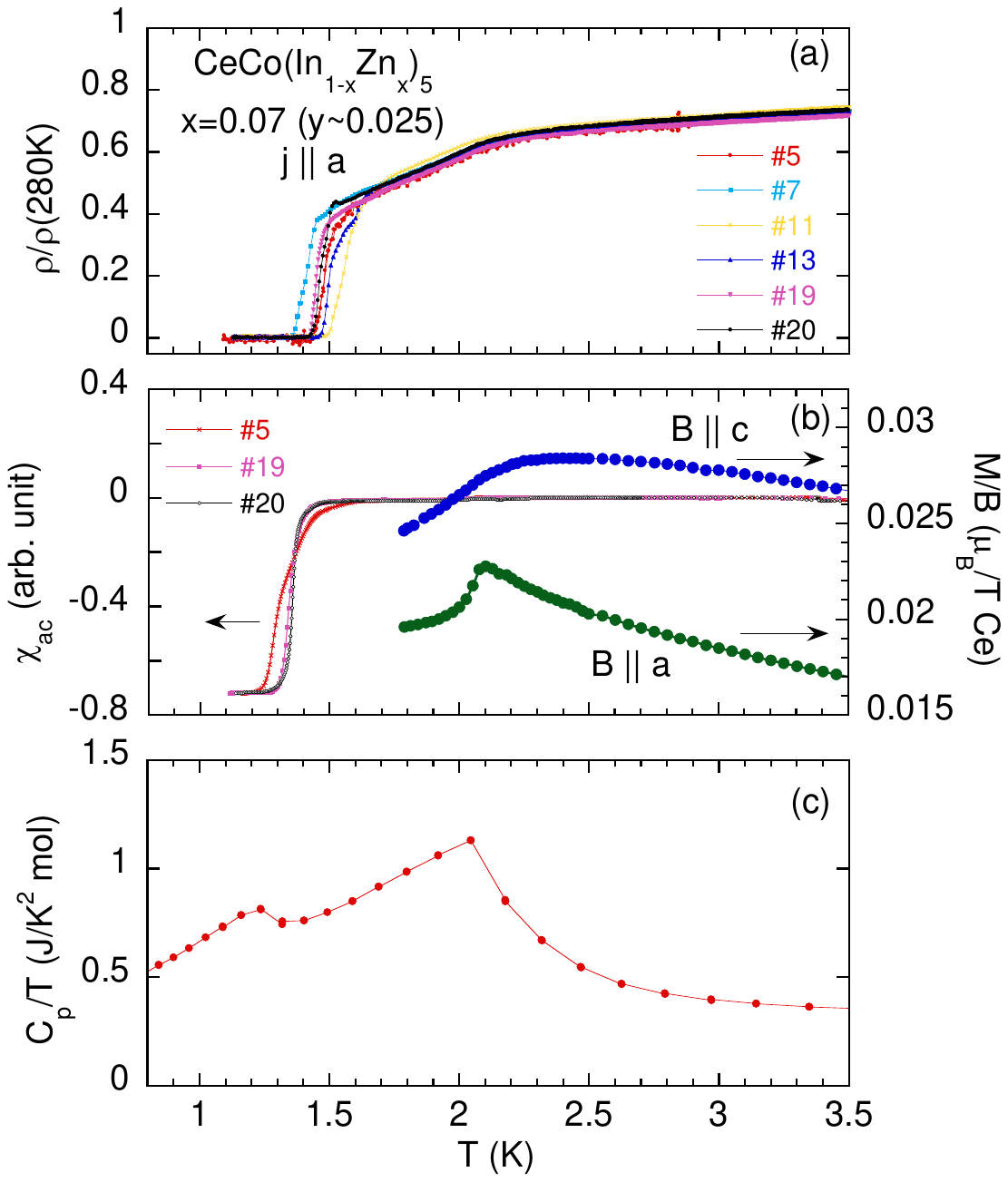}
\end{center}
  \caption{
(Color online)  Temperature variations of (a) the $a$-axis electrical resistivity $\rho$ normalized by the magnitude at 280 K,  (b) the $a$- and $c$-axis magnetization divided by magnetic field of $B=0.1\ {\rm T}$ and the $a$-axis ac susceptibility $\chi_{ac}$, and (c) the specific heat divided by temperature for CeCo(In$_{0.93}$Zn$_{0.07}$)$_5$. In (a) and (b), the $\rho$ and $\chi_{ac}$ data taken from various samples are shown for comparison. 
}
\end{figure}

Specific heat $C_p$ was measured with a thermal relaxation method, and electrical resistivity $\rho$ measurements were performed with a standard four-wire technique. Both the experiments were done in temperatures down to 0.5 K and external magnetic field $B\ (=\mu_0H)$ ($\mu_0$: vacuum permeability) of 0-14 T using commercial measurement units (PPMS, Quantum Design) and hand-made equipment. In the $\rho$ measurements, the magnetic field is applied perpendicular to the sample current. Magnetization $M$ was measured using a capacitively detected Faraday-force magnetometer\cite{rf:Sakakibara94} installed in a $^3$He cryostat (Heliox, Oxford Instruments), in which the measured temperature and magnetic-field ranges were 0.27-3 K and 0-14 T, respectively. A commercial SQUID magnetometer (MPMS, Quantum Design) was also utilized for the magnetization experiments in the temperature range of 1.8-300 K and fields up to 7 T. To check the SC transition of the samples, ac susceptibility $\chi_{\rm ac}$ was measured between 1.1 and 4 K using a conventional Hartshorn-bridge method. A frequency and an amplitude of applied ac field were selected to be $\sim 0.05\ {\rm mT}$ and 180 Hz. 

\section{Results}
\subsection{Low-temperature upper critical fields}
Figure 2 shows field variations of isothermal magnetization $M(B)$ at 0.27 K, which are measured by applying $B$ along the $a$ and $c$ axes. The SC order is recognized from the hysteretic behavior in $M(B)$ measured under increasing and decreasing field variations, and an enlarged hysteresis loops by increasing $x$ would be ascribed to a disorder effect induced by doping. The SC upper critical fields $H_{c2}$ are determined by the closing of the hysteresis loop in $M(B)$. At $x=0.025$, a strong peak effect seen in $M(B)$ just below $H_{c2}$ obscures the thermodynamic features of $M(B)$ which ought to be observed at the suppression of the SC order parameter for both field directions. This peak effect also manifests itself in the field variations of magnetic torque roughly estimated by the capacitance data of the magnetometer taken without applying the gradient field [inset of Fig. 2(a)]. On the other hand, $M(B)$ for $x \ge 0.05$ ($y\ge 0.019$) continuously changes at around $H_{c2}$, suggesting that the breakdown of the SC order occurs as a second-order transition at $H_{c2}$. This is in contrast with the situation in pure CeCoIn$_5$, in which the transition from the SC to the normal states generated by $B$ is revealed to be of the first order below 0.7 K.\cite{rf:Izawa2001,rf:Tayama2002,rf:Ikeda2001,rf:Bianchi2002} Similar change in the order of the phase transition has also been reported in low Cd concentrations of the mixed alloys CeCo(In,Cd)$_5$, and is argued on the basis of the impurity effect.\cite{rf:Tokiwa2008} The $c$-axis $M(B)$ curves for $x \ge 0.05$ ($y\ge 0.019$) show additional kink- or shoulder-like behavior at $\sim 5\ {\rm T}$ ($=B_M$) [Fig.\ 2(b) and its inset]. This anomaly is distinguished from the SC boundary, and attributed to a disappearance of the AFM order, since $B_M$ decreases with increasing temperature and then becomes zero at $T_N$.\cite{rf:Yoko2014} For $B\,||\, a$, by contrast, $M(B)$ do not exhibit such an anomaly in the field range presently investigated, reflecting the anisotropy of the AFM stability against magnetic fields as discussed later.
\begin{figure}[tbp]
\begin{center}
\includegraphics[bb=8 238 534 807,keepaspectratio,width=0.45\textwidth]{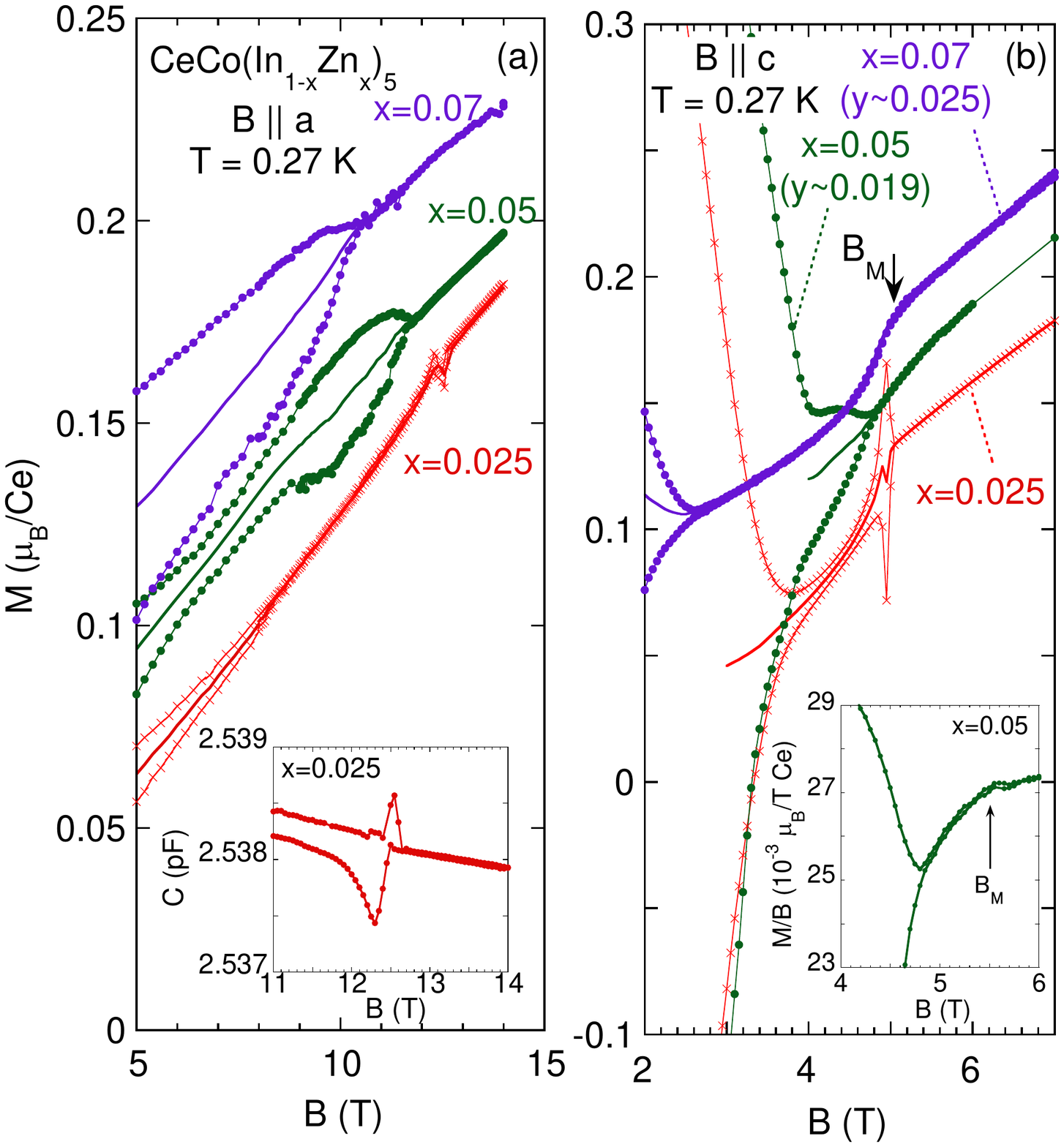}
\end{center}
  \caption{
(Color online)  Isothermal magnetization curves at 0.27 K for CeCo(In$_{1-x}$Zn$_{x}$)$_5$ ($x=0.025$, 0.05 and 0.07), measured under $B$ with the directions of (a) $B\,||\,a$ and (b) $B\,||\,c$. Note that the offsets of 0.025 $\mu_{\rm B}/{\rm Ce}$ and 0.05 $\mu_{\rm B}/{\rm Ce}$ are added to the magnetization data at $x=0.05$ ($y\sim 0.019$) and 0.07 ($y\sim 0.025$) for clarity. Solid lines are the average of the magnetization curves between increasing and decreasing field variations. Shown in the inset of (a) are the raw capacitance data ($B\,||\,a$) for $x=0.025$ measured without applying a gradient field. The enlargement of the $c$-axis $M/B$ curve for $x=0.05$ is shown in the inset of (b).
}
\end{figure}

The present magnetization measurements demonstrate that for both the $a$ and $c$ directions, $H_{c2}$ at low temperatures is almost invariant under the Zn doping at least for $x$ less than 0.05. In Fig.\ 3(a), we plot the $x$ variations of $H_{c2}$ at 0.27 K for both directions, $H_{c2}^a$ and $H_{c2}^c$, and compare with the zero-field $x-T$ phase diagram [Fig.\ 3(b)].\cite{rf:Yoko2014} Both $H_{c2}^a$ and $H_{c2}^c$ slightly increase with increasing $x$, from the values at $x=0$ ($\mu_0H_{c2}^a=11.5\ {\rm T}$ and $\mu_0H_{c2}^c=4.9\ {\rm T}$) \cite{rf:Tayama2002} and then approach the maximum at $x=0.025$. They turn to decrease on further doping, but even at $x=0.05$ ($y\sim 0.019$), they still take the values as large as those at $x=0$, although $T_c\ (=1.8\ {\rm K})$ at $x=0.05$ is reduced to 80\% of that for pure CeCoIn$_5$. These features seen in the $x$ variations of $H_{c2}$ are in stark contrast to the monotonic reduction of $T_c$ by increasing $x$ at zero field. The increase of $H_{c2}$ is also observed in CeCo(In,Cd)$_5$ and CeCo(In,Hg)$_5$, \cite{rf:Tokiwa2010} indicative of the common mechanism involved in these doped alloys. At $x=0.07$ ($y\sim 0.025$), however, the reduction of $H_{c2}$ becomes anisotropic; the reduction ratios $H_{c2}(x=0.07)/H_{c2}(x=0)$ at 0.27 K are 0.9 for $B\, ||\,a$ and 0.6 for $B\, ||\,c$. They should be compared with $T_c(x=0.07)/T_c(x=0)$ of about 0.6. Accordingly, a relationship between the anisotropic decrease of $H_{c2}$ and the AFM order is inferred since a clear AFM transition occurs in this $x$ range. 
\begin{figure}[tbp]
\begin{center}
\includegraphics[bb=38 498 423 816,keepaspectratio,width=0.45\textwidth]{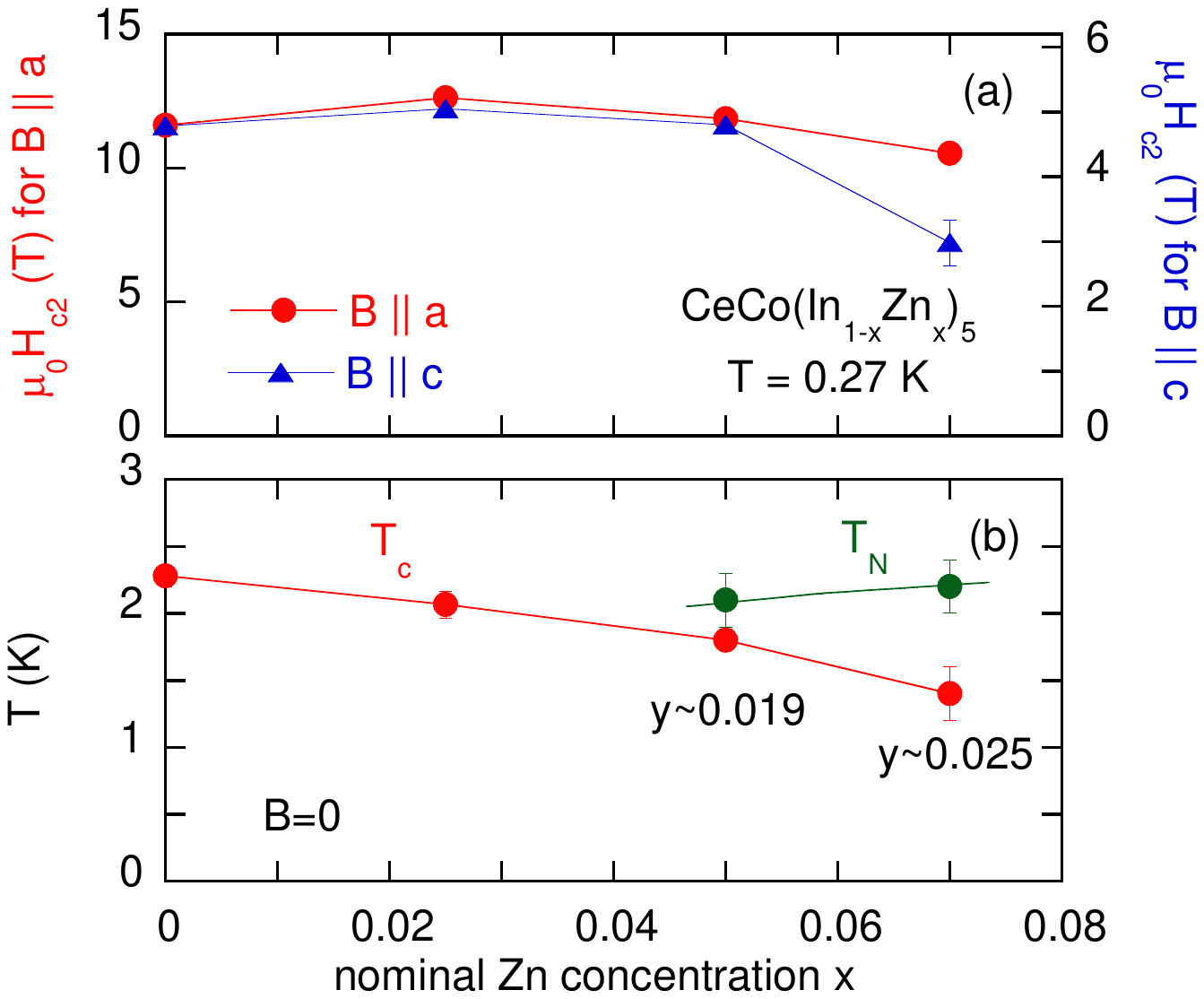}
\end{center}
  \caption{
(Color online)  (a) The $a$- and $c$-axis upper critical fields $H_{c2}$ at 0.27 K for CeCo(In$_{1-x}$Zn$_{x}$)$_5$ plotted as a function of a nominal Zn concentration $x$. (b) The zero-field $x-T$ phase diagram for CeCo(In$_{1-x}$Zn$_{x}$)$_5$ obtained from the temperature variations of the ac susceptibility, magnetization, electrical resistivity and specific heat.\cite{rf:Yoko2014}
}
\end{figure}

\subsection{Superconducting and antiferromagnetic orders under a magnetic field}
The characteristics of the SC and AFM orders can also be explored by means of thermal and transport experiments under magnetic fields. Displayed in Fig.\ 4 are the temperature variations of the electrical resistivity $\rho(T)$ near $T_c$ for $x=0.07$ ($y\sim 0.025$), measured by applying $B$ along the $a$ and $c$ axes. The SC transition is indicated by a sharp drop of $\rho(T)$ to zero. The transition temperature is lowered by increasing $B$ for both field directions. $\rho(T)$ at 10 and 11 T for $B\,||\,a$ exhibits a partial drop without reaching zero resistivity in the temperature range of the present experiments, and the onset of the transition completely disappears above $12\ {\rm T}$. This is roughly consistent with the results of the magnetization measurement, in which $\mu_0H_{c2}$ at 0.27 K is estimated to be 10.6 T. Similar features, along with the correspondence to the $M(B)$ behavior, are also found in the $\rho(T)$ data obtained for $B\,||\,c$. However, in this field direction,  the SC transition width becomes significant even in the low $B$ range, $\sim 0.4\ {\rm K}$ at 2 T, in contrast with the case for $B\,||\,a$ in which a sharp drop of $\rho$ at $T_c$ continues up to $\sim 10\ {\rm T}$.
\begin{figure}[tbp]
\begin{center}
\includegraphics[bb=21 386 517 778,keepaspectratio,width=0.48\textwidth]{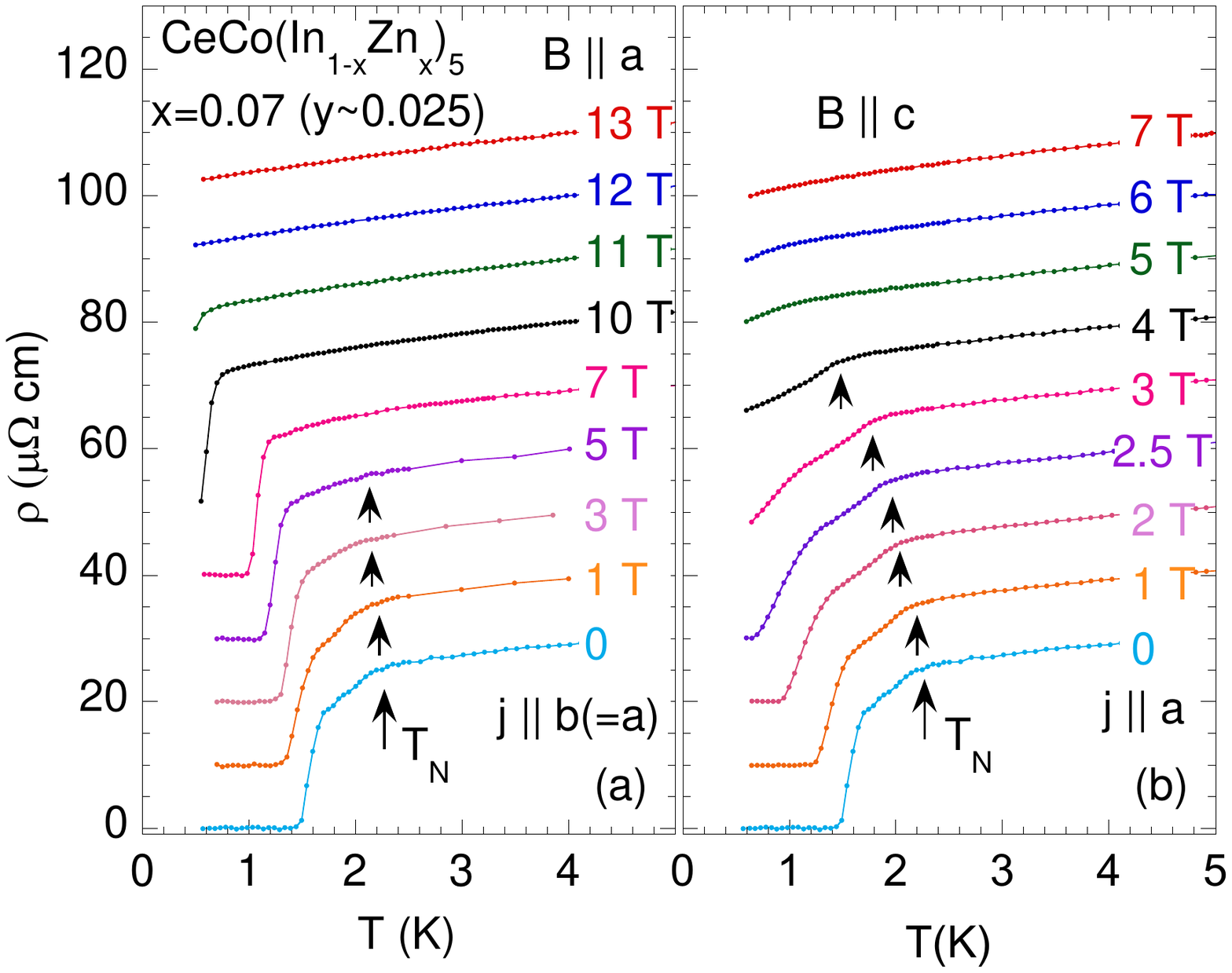}
\end{center}
  \caption{
(Color online)  Temperature variations of the $a$-axis electrical resistivity $\rho(T)$ for CeCo(In$_{0.93}$Zn$_{0.07}$)$_5$, measured under various $B$ with the directions of (a) $B\,||\,a$ and (b) $B\,||\,c$. Note that the magnetic field is applied perpendicular to the sample current $j$ in both (a) and (b). The $\rho(T)$ data taken under the different field strengths are vertically shifted in a step of 10 $\mu\Omega\ {\rm cm}$ for clarity. The arrows indicate the positions of the kink originating from the AFM order.
}
\end{figure}

The feature in $\rho(T)$ associated with the AFM transition is very much dependent on the field direction. Applying $B$ along the $a$ axis rapidly weakens the kink at $T_N$, although there is only a slight decrease of $T_N$ from 2.2 K ($B=0$) to $\sim 2.1$ K (5 T). The kink finally fades out above 7 T. For $B\,||\,c$, on the other hand, a rather clear kink anomaly is observed at $T_N$ up to 4 T even though $T_N$ is reduced to 1.5 K. The anomaly can still be seen at 6 T as a weak convex curvature in $\rho(T)$ near the base temperature. A similar trend can also be found in the $M(B)$ curves at low temperatures; $M(B)$ for the $a$ axis shows no anomaly that can be ascribed to the AFM to a paramagnetic transition within the field range of the present measurements, whereas $M(B)$ for the $c$ axis exhibits the shoulder-like anomaly at $B_M\sim 5\ {\rm T}$ (Fig. 2). We have previously shown that $\rho(T)$ in the paramagnetic region above $T_N$ for Zn-doped samples is governed by the non-Fermi-liquid behavior characterized by a $T^n$ dependence in $\rho(T)$ with $n$ ranging from $\sim 0.5$ to 1.\cite{rf:Yoko2014} At $x=0.07$ ($y\sim 0.025$), the exponent $n$ is found to be still much smaller than the value expected from the Fermi-liquid state ($n=2$) in the whole $B$ range studied, $B \le 14\ {\rm T}$ ($B\,||\,a$) and $B \le 7\ {\rm T}$ ($B\,||\,c$), indicating that the origin of quantum critical fluctuations and concomitant scattering of conduction electrons is not significantly affected by the application of a magnetic field.

Temperature variation of the specific heat $C_p$ for $x=0.07$ ($y\sim 0.025$) also demonstrates that the SC and AFM transitions differently respond to the field, depending on the field directions. In Fig.\ 5, we plot the specific heat divided by temperature, $C/T\equiv (C_p-C_{\rm nucl})/T$, obtained under various fields along the $a$ and $c$ axes, where the contribution of the nuclear Schottky anomaly $C_{\rm nucl}$, observable only at $T\sim 0.5\ {\rm K}$ and $B\sim14\ {\rm T}$, is carefully subtracted by calculations based on the natural abundance of nuclear spins in the samples. $C/T$ at $B=0$ exhibits a small but clear jump at $T_c=1.3\ {\rm K}$, reflecting the SC transition. It remains sharp up to 10 T for $B\,||\,a$, while $T_c$ is continuously reduced toward zero with increasing $B$. However, the large jump associated with the AFM transition at $T_N=2.2\ {\rm K}$ is rapidly damped by applying $B$ along the $a$ axis. It becomes indiscernible above 5 T, leaving a shoulder-like weak anomaly at $\sim 2\ {\rm K}$. In contrast, we find quite opposite evolution of the broadening of the $C/T$ jumps when $B$ is applied along the $c$ axis: the AFM transition accompanies a clear and large jump of $C/T$ in the $B$ range up to 4 T ($\sim B_M$), whereas the jump at $T_c$ is rapidly broadened by a small field of $\sim 1\ {\rm T}$. The broadening of the SC transition temperature, estimated by the specific heat jump, amounts to $\sim 0.3\ {\rm K}$ at $B=1.5\ {\rm T}$, which is comparable to the transition width determined from $\rho(T)$ using the same sample. Note that for $B\,||\,c$, $C/T$ near $B_M$ is strongly enhanced at low temperatures. This feature can still be seen at $B=7\ {\rm T}$; $C/T$ shows a tendency of saturation to a large value of 0.8 J/K$^2$ mol. The $C/T$ constant behavior becomes invisible down to 0.5 K at $B=9\ {\rm T}$. Interestingly, the $C/T$ data at $B=9\ {\rm T}$ obey a $-\ln T$ dependence below $\sim 1.5\ {\rm K}$ [inset of Fig.\ 5(b)], suggesting that the quantum critical fluctuations develop at this field. It is likely that the short-ranged AFM correlation emerging above $B_M$ is an origin for the $C/T$ saturation and the weak convex behavior in $\rho(T)$, and this effect seems to make the fluctuation region stay away from $\sim B_M$. In addition, the evolution of the AFM quantum critical fluctuations is considered to depend on the field direction because the strong enhancement of $C/T$ is never detected for $B$ along the $a$ axis. At $B=14\ {\rm T}$, $C/T$ for both field directions eventually show a weak $T$-linear increase with decreasing temperature.
\begin{figure}[tbp]
\begin{center}
\includegraphics[bb=7 208 531 819,keepaspectratio,width=0.48\textwidth]{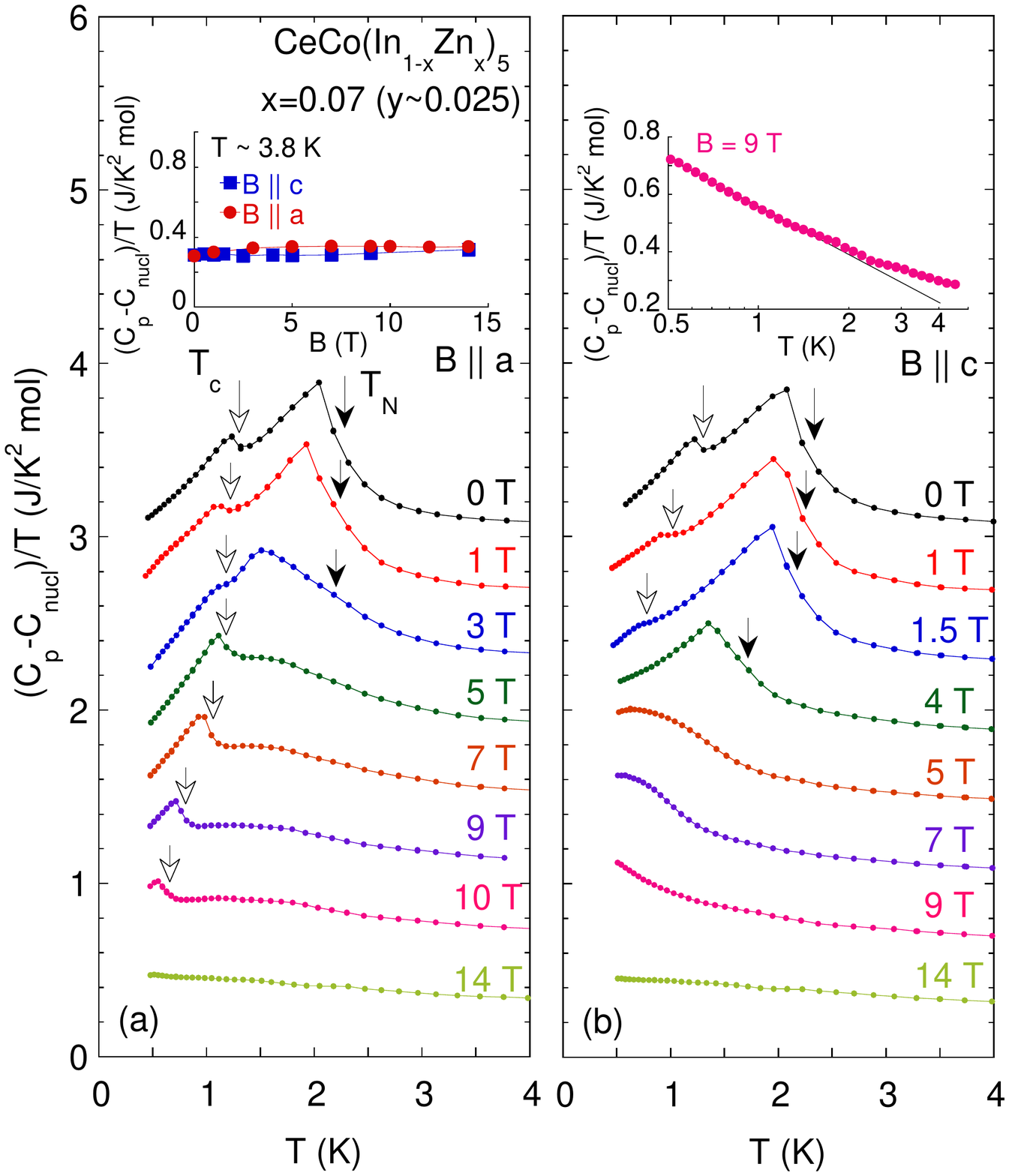}
\end{center}
  \caption{
(Color online) Temperature variations of the specific heat divided by temperature, $C/T\equiv (C_p-C_{\rm nucl})/T$, for CeCo(In$_{0.93}$Zn$_{0.07}$)$_5$, measured under various $B$ with the directions of (a) $B\,||\,a$ and (b) $B\,||\,c$. The nuclear Schottky contribution $C_{\rm nucl}$ is subtracted using the calculations based on the natural abundance of nuclear spins in the samples. The baseline of the $C/T$ values obtained at each field is shifted in a step of 0.4 ${\rm J/K^2\ mol}$ from that at 14 T for clarity. The open and closed arrows indicate $T_c$ and $T_N$, respectively. $C/T$ versus $B$ at $T\sim 3.8\ {\rm K}$ is shown in the inset of (a). The inset of (b) is the $\log T$ plot of $C/T$ at 9 T for $B\,||\,c$.
}
\end{figure}

The evolution of the NFL divergence in $C/T$ is quite in contrast with the nondiverging feature, such as the Fermi-liquid behavior, observed in $C/T$ of CeCo(In$_{1-x}$Cd$_x$)$_5$, although the electrical resistivity for both alloys indicates no clear signature of the $T^2$ dependence around the AFM order.\cite{rf:Seo2014,rf:Nair2010} A possible reason for this discrepancy between Zn- and Cd- doped alloys arises from the difference of the order of the AFM transition under the magnetic field. If the AFM to paramagnetic transition in the high-field and low-temperature region is of the second order, the AFM fluctuation would be enhanced above $B_M$, yielding the NFL behavior in the temperature variations of $C/T$. By contrast, the AFM fluctuation may be suppressed above $B_M$ when the transition is of the first order. The difference of the order of the AFM transition might also be connected with the broadening of the AFM transition for $B\,||\, a$ in CeCo(In$_{1-x}$Zn$_x$)$_5$. It is supposed that the first-order AFM transition at low fields yields the broadening of the transition for $B\,||\, a$, but this transition progressively changes into a second-order one at high fields for $B\,||\, c$. At present, our investigations on the dc magnetization for increasing and decreasing magnetic fields and the temperature variations of the electrical resistivity reveal no indication of the hysteresis associated with the AFM transition within the experimental accuracy. We consider that the direct observation of the AFM order by means of microscopic probes would be useful to verify this possibility.

Displayed in Fig. 6 are the temperature variations of the entropy divided by the gas constant, $S/R$, for $x=0.07$ ($y\sim 0.025$), derived by integrating the $C/T$ data above 0.5 K. Since the magnitude of $C/T$ at $\sim 3.8\ {\rm K}$ ($> T_c$ and $T_N$) is field independent for both the $a$ and $c$ directions [the inset of Fig. 5(a)], it is considered that the entropy release associated with the low-temperature phase transitions is completed at this temperature for all fields studied. Therefore, we set the initial values of $S/R$ at 0.5 K in Fig. 6 so that the $S/R$ values at 3.8 K for different fields become equal to each other. For $B\,||\,a$, the onset always starts at $\sim 2.2\ {\rm K}$ regardless of the strength of $B$, although the reduction of $S/R$ is broadened with increasing $B$ and becomes unclear at $B=14\ {\rm T}$. In contrast, the onset for $B\,||\,c$ rapidly decreases in temperature from $\sim 2.2\ {\rm K}$ ($B=0$) to 1.5 K ($B=9\ {\rm T}$). These features are consistent with the observations in the $\rho$ and $C/T$ measurements. Note that the low-temperature residual entropy, estimated from the difference $[S(B)-S(B=0)]/R$ at 0.5 K in Fig. 6, has the same magnitude of $\sim 0.07$ at $B=14\ {\rm T}$ for both  directions. Temperature dependence of $S/R$ at this field is nearly the same for the two directions. These correspondences strongly suggest that the AFM order is replaced by the paramagnetic spin polarized state at $B=14\ {\rm T}$ for $B\,||\,a$ as well as $B\,||\,c$, although the AFM phase transition is smeared at high fields in the former direction. 
\begin{figure}[bp]
\begin{center}
\includegraphics[bb=29 335 457 804,keepaspectratio,width=0.43\textwidth]{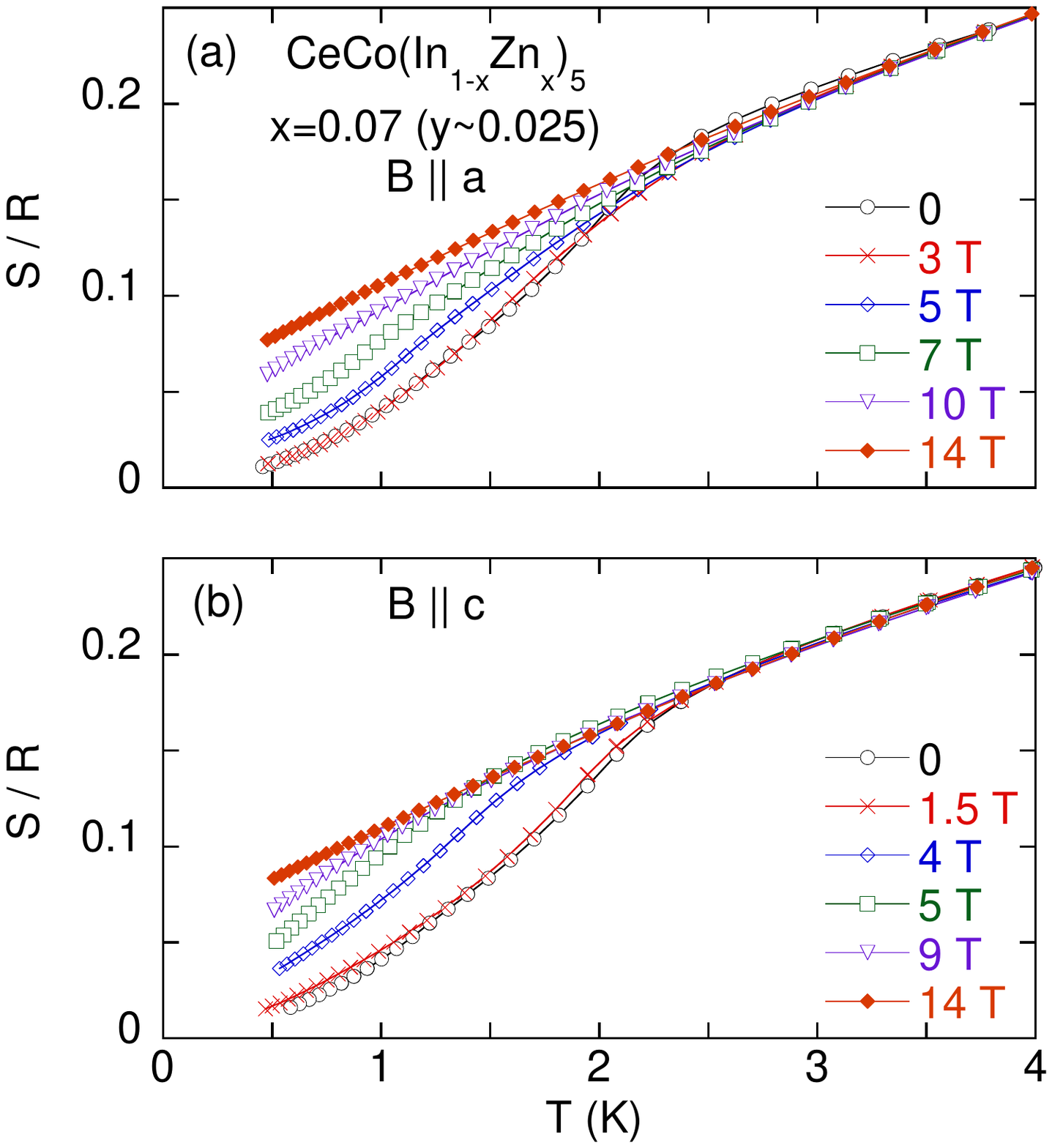}
\end{center}
  \caption{
(Color online) The entropy change divided by the gas constant, $S/R$, for CeCo(In$_{0.93}$Zn$_{0.07}$)$_5$ with (a) $B\,||\,a$ and (b) $B\,||\,c$, plotted as a function of temperature. The entropy change is evaluated by integrating $(C_p-C_{\rm nucl})/T$ above 0.5 K. The initial values of $S/R$ at 0.5 K are modified so that the $S/R$ values at 3.8 K coincide to each other. 
}
\end{figure}

Furthermore, a possible relationship in the quantum critical fluctuation is found between pure and Zn-doped CeCoIn$_5$ from a comparison of the entropy changes associated with the non-Fermi liquid anomaly. Namely, the magnitude and temperature dependence of $S/R$ for $B > B_M$ ($x=0.07$) and $B > \mu_0H_{c2}$ ($x=0$) almost coincide with each other,\cite{rf:Bauer2011} indicating that the low-energy degrees of freedom relevant to the quantum critical fluctuation are basically unchanged through the Zn substitution.

The magnetic-field versus temperature ($B-T$) phase diagrams resulting from the specific-heat, electrical-resistivity, and isothermal-magnetization measurements are summarized in Fig.\ 7. The phase boundaries derived from these measurements are consistent with each other except the slight discrepancy ($\sim 0.1\ {\rm K}$) in the SC boundaries estimated from the electrical resistivity and the other measurements. This difference can also be recognized at zero field (Fig.\ 1). A similar trend, a discrepancy in the definition of $T_c$ among different measurement techniques, has often been reported in the other heavy-fermion superconductors having a sensitivity to sample quality, such as UCoGe.\cite{rf:Huy2007} We thus consider this caused by the effect of disorder in the samples. Indeed, we find that the discrepancy in the SC transition temperatures appears only for the doped samples with $x=0.07$ ($y\sim 0.025$). 
\begin{figure}[tbp]
\begin{center}
\includegraphics[bb=26 226 457 800,keepaspectratio,width=0.43\textwidth]{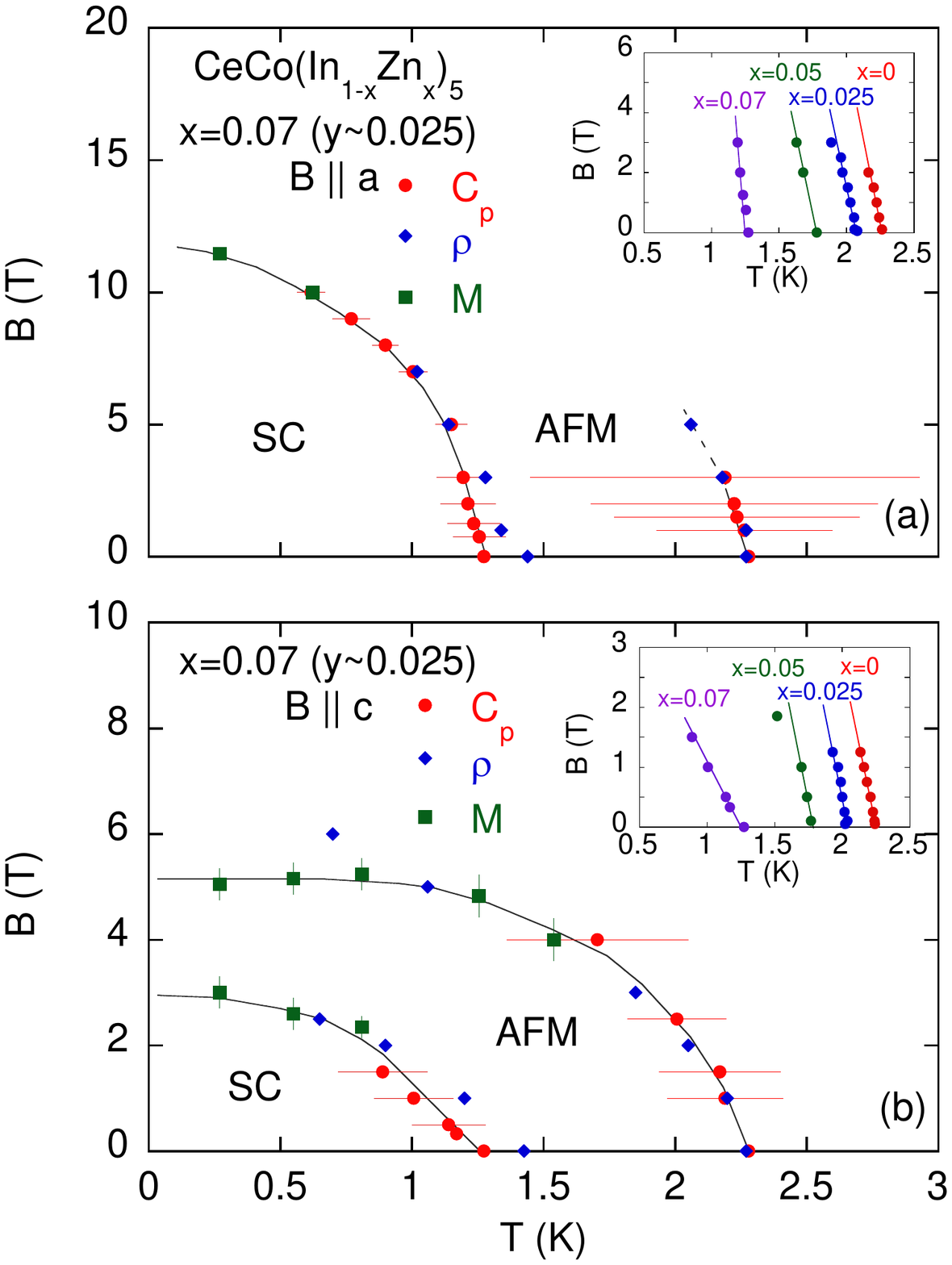}
\end{center}
  \caption{
(Color online)  Low-temperature $B-T$ phase diagram of CeCo(In$_{1-x}$Zn$_{x}$)$_5$ ($x=0.07$, $y\sim 0.025$) with (a) $B\,||\,a$ and (b) $B\,||\,c$, derived from the anomalies seen in the specific heat, electrical resistivity and isothermal magnetization. The horizontal error bars indicate the broadening of the AFM and SC transitions estimated from the derivative of the heat capacity as a function of temperature. The insets show the slope of $\mu_0H_{c2}(T)$ at $\sim T_c$ for $x=0$, 0.025, 0.05, and 0.07 derived from the specific heat and the magnetization data. The lines on the phase boundaries are guides to the eye.
}
\end{figure}

The obtained $B-T$ phase diagram for $x=0.07$ ($y\sim 0.025$) indicates that the AFM order is strongly anisotropic. The AFM phase boundary definitely surrounds the SC phase region for $B\,||\,c$, but this feature is less clear in the high-field region for $B\,||\,a$. This is primarily beause the $B$-induced broadenings of the AFM boundary, which are indicated as the horizontal bars in Fig.\ 7, are much larger for $B\,||\,a$ than for $B\,||\,c$. These differences might arise from a specific AFM structure and spin polarizations induced in the Zn-doped compounds. In fact, the isostructural Ce115 compounds have a rich variety of AFM structures as will be discussed in the following section. 

It is interesting that a stark difference in the $B-T$ phase diagram of the AFM order can be found between Zn- and Cd-doped CeCoIn$_5$, since in CeCo(In$_{1-x}$Cd$_{x}$)$_5$, the AFM transition temperature is definitely determined in all the field range for $B\perp c$ as well as $B\,||\,c$, and the AFM phase boundary simply covers the SC phase for both the field directions.\cite{rf:Nair2010} This discrepancy, along with the field-induced evolution of the NFL behavior seen only in CeCo(In$_{1-x}$Zn$_{x}$)$_5$, may arise from the difference in the AFM structures or the order of the AFM transition under the magnetic field.

\section{Discussion}
We previously presented that the substitution of Zn for In leads to little change of the $c$-axis $H_{c2}$ while it monotonically reduces $T_c$.\cite{rf:Yoko2014} The present investigations have further clarified that such a robust $x$ dependence of $H_{c2}$ appears in both $a$ and $c$ directions for $x \le 0.05$ ($y\le 0.019$), and then the anisotropic reduction of $H_{c2}$ becomes significant for $x=0.07$ ($y\sim 0.025$). The breakdown of the SC state by magnetic field is usually governed by the orbital-pair-breaking mechanism in type-II superconductors. It is known in a simple situation that the orbital-limited critical field at zero temperature $H_{c2}^{\rm orb}(0)$ is connected with the initial slope of $H_{c2}$ at $T_c$ via the relation $\mu_0H_{c2}^{\rm orb}(0) \sim -0.7\, T_c\, (\mu_0dH_{c2}/dT)_{T_c}$,\cite{rf:Werthammer66}, and $-(\mu_0dH_{c2}/dT)_{T_c}$ is roughly in proportion to the effective mass of conduction electrons. One may thus expect that the robust $x$ variation of $H_{c2}$ found in CeCo(In$_{1-x}$Zn$_{x}$)$_5$ is attributed to the mass enhancement of the heavy-fermion quasiparticles counterbalancing the reduction of $T_c$. However, such an interpretation is not be applicable to these alloys, as described below. Here, we estimate the $H_{c2}^{\rm orb}(0)$ values from the initial slope of $H_{c2}$. Displayed in the insets of Fig. 7 are the temperature variations of the upper critical field $H_{c2}(T)$ around $T_c$, which are derived from the specific-heat and magnetization measurements. It should be taken into account that the simple situation leading to the above formula for $H_{c2}^{\rm orb}(0)$ might not be realized in CeCo(In$_{1-x}$Zn$_{x}$)$_5$ with $x \sim 0.07$, since $H_{c2}(T)$ is expected to be influenced by the AFM order when $T_c$ becomes significantly smaller than $T_N$. At $x=0.07$, in fact, the field-induced broadening of the SC phase boundary occurs even in the low $B$ range for $B\,||\,c$, while being absent for $B\,||\,a$, and hampers the estimation of the initial slope of the $c$-axis $H_{c2}$. For $x < 0.07$, the initial slopes of $H_{c2}$ for both field directions are roughly independent of $x$, yielding $\mu_0H_{c2}^{\rm orb}(0)$ ranging from 25 to 31 T for $B\, ||\, a$, and from 16 to 20 T for $B\, ||\, c$. These values are still much larger than the actual $\mu_0H_{c2}$ observed in the present investigations, suggesting that the spin-pair-breaking mechanism governs the breakdown of the SC order at $H_{c2}$ even in the Zn-doped samples, and hence, the orbital-pair-breaking effect cannot simply account for the robust $x$ variation of $H_{c2}$. As for the situation in the $c$-axis $H_{c2}$ for $x=0.07$, the estimated $\mu_0H_{c2}^{\rm orb}(0)$ falls in the range of 3-6 T though the ambiguity in the experimental determination of $H_{c2}(T)$ is fairly large. This range is comparable to the observed $\mu_0H_{c2}$, implying that the orbital-pair-breaking mechanism becomes effective only for $B\,||\,c$ at $x=0.07$. 

When the strong Pauli paramagnetic effect governs the breakdown of superconductivity by a magnetic field, the Pauli-limited field $H_P$ is determined through the competition between the Zeeman energy of conduction electrons and the SC condensation energy as follows:
\begin{eqnarray}
\frac{1}{2}\mu_0\chi_p H_P^2=\frac{1}{2}D(\varepsilon_F)\Delta^2,
\end{eqnarray} 
where $\chi_p$, $D(\varepsilon_F)$ and $\Delta$ denote the paramagnetic spin susceptibility, density of states at the Fermi level, and the SC gap, respectively. Assuming that the conditions $H_{c2}\sim H_P$ and $H_{c2}^{\rm orb} \gg H_P$ hold in CeCo(In$_{1-x}$Zn$_{x}$)$_5$, the robust $x$ variation of $H_{c2}$ would be attributed to the suppressions of $\chi_p$ as well as $\Delta$ with doping Zn. Note that the AFM order emerges for $x\ge 0.05$ ($y\ge 0.019$), and $T_c$ is monotonically reduced with increasing $x$.\cite{rf:Yoko2014} In this situation, the AFM correlations are expected to develop even at low Zn concentrations, and will suppress $\chi_p$ at low temperatures, although this effect might not be detected by the magnetization measurement because of a shielding by the SC order. At the same time, it is obvious that $\Delta$ is also reduced by doping Zn in connection with a drop in $T_c$. The reductions of both quantities make $H_P$ insensitive to $x$ in Eq.\ (1); in particular, the Pauli paramagnetic suppression is relaxed by the AFM correlations in Zn-doped CeCoIn$_5$. In this context, the situation is more complicated at $x\sim 0.07$ ($y\sim 0.025$) because the long-range AFM order is well stabilized and the magnetic anisotropy becomes salient. There the AFM order parameter should be coupled with both the Zeeman and the SC condensation energies in Eq.\ (1). If this is the case, the observed difference in the $H_{c2}$ reduction between the $a$- and $c$-axis magnetic fields at $x=0.07$ would be related to the spin structure and polarization in the AFM order as well as the competition between spin- and orbital-pair-breaking effects.  

The present investigations have also revealed the anisotropic response of the AFM order to the magnetic field. For $B\,||\,c$, a continuous suppression of the AFM order can be easily traced down to $T_N \sim 0$ by the sharp anomalies in the thermodynamic quantities at the transition temperature. By contrast, the signatures of the AFM phase transition at $T_N$ are rapidly smeared by increasing $B$ along the $a$ axis, making it difficult to detect the reduction of $T_N$. The field variations of the magnetization and entropy suggest that the AFM spin modulation rather abruptly changes to the uniform polarized states at $B_M$ for $B\,||\,c$, whereas a uniformly polarized spin component is coexisting with the AFM modulation and gradually evolves for $B\,||\,a$. Furthermore, at $x=0.07$ ($y\sim 0.025$) the Pauli paramagnetic effect is still dominant on $H_{c2}$ for $B\,||\,a$, whereas it seems to be relatively weak for $B\,||\,c$. This difference in the Pauli paramagnetic effect would be explained by the anisotropy of the magnetic susceptibility in the AFM phase; the reduction of the $a$-axis susceptibility due to the AFM ordering is expected to be significantly smaller than that of the $c$-axis susceptibility. All of the above features suggest that the anisotropy of the spin polarization is not so strong in the $c$ plane, resulting in a gradual polarization of the AFM spins toward the $B$ direction. Indeed, it has been revealed that the isostructural Ce115 compounds have a rich variety of magnetic structures characterized by the wave vector of $q=(1/2,1/2,\delta)$ and the spiral spin polarization in the $c$ plane.\cite{rf:Yoko2006,rf:Ohira-Kawamura2007,rf:Yoko2008,rf:Nicklas2007,rf:Bao2000,rf:Raymond2007,rf:Raymond2014-2} CeRhIn$_5$ shows the AFM transition below $T_N=3.8\ {\rm K}$, whose structure has been proposed to be helical along the $c$ axis with an incommensurate propagation vector of $\delta=0.297$.\cite{rf:Hegger2000,rf:Bao2000} The AFM propagation vector changes into a commensurate modulation of $\delta=1/4$ by applying $B$ along the $c$ plane.\cite{rf:Raymond2007} The AFM order with a further reduced $\delta$ value toward zero can be generated by doping Sn.\cite{rf:Raymond2014-2} On the other hand, the mixed compounds CeRh$_{1-x}$Co$_x$In$_5$ exhibit possible multiple AFM structures with $\delta=0.42$ and 1/2 as $x$ is increased.\cite{rf:Yoko2008} These instabilities in the spiral AFM modulations could also be involved in CeCo(In$_{1-x}$Zn$_{x}$)$_5$, and might be responsible for the absence of the clear AFM phase boundary for $B\,||\,a$. 

\section{Conclusion}
We presented thermal, magnetic, and transport properties of the SC and AFM orders under the magnetic field for CeCo(In$_{1-x}$Zn$_x$)$_5$ with $x \le 0.07$ ($y\le 0.025$). It is found that the substitution of Zn for In leads to little change of $H_{c2}$ for both the $a$ and $c$ axes, while it monotonically reduces $T_c$. $H_{c2}$ for these directions show slight increase with increasing $x$ and then have the maximum at $x=0.025$. They are reduced by further doping, but even at $x=0.05$ ($y\sim 0.019$), they still take the values as large as those at $x=0$, though $T_c$ is reduced to 80\% of that for pure CeCoIn$_5$. At $x=0.07$ ($y\sim 0.025$), on the other hand, the anisotropic reduction of $H_{c2}$ becomes prominent. We suggest from the analysis based on the initial slopes of $H_{c2}$ that the strong Pauli paramagnetic effect governs the pair-breaking mechanism at $H_{c2}$ in CeCo(In$_{1-x}$Zn$_x$)$_5$. In this context, it is considered that the nearly invariant $H_{c2}$ with doping Zn is attributed to the combination of the reduced spin susceptibility due to the AFM correlation and the suppression of the SC condensation energy. 

It is also found that the AFM order shows the anisotropic response to the magnetic field at $x=0.07$ ($y\sim 0.025$). For $B\,||\,c$, a continuous suppression of the AFM order can be recognized down to $T_N\sim 0$ by the sharp anomalies in the thermodynamic quantities at the transition temperature. However, the signatures on the AFM transition in these quantities are rapidly damped and then become unclear by increasing $B$ along the $a$ axis. In particular, at 0.27 K, $M(B)$ for the $a$ axis shows no clear indication of the transition from the AFM to the spin polarized states up to $B=14\ {\rm T}$, whereas $M(B)$ for the $c$ axis exhibits a shoulder-like anomaly at $B_M\sim 5\ {\rm T}$. In addition, the magnitude of the low-temperature entropy and its temperature dependence are nearly the same for two directions at $B = 14\ {\rm T}$. These experimental results indicate that the AFM spin modulation rather abruptly changes to the uniform polarized states at $B_M$ for $B\,||\,c$, whereas a uniformly polarized spin component coexists with the AFM modulation and gradually evolves for $B\,||\,a$. We suggest that this difference might arise from instabilities in the spiral AFM modulations along the $c$ axis possibly involved in the Zn-doped compounds.

\begin{acknowledgments}
M.Y. is grateful to I. Kawasaki, S. Nakamura and S. Kittaka for experimental supports and to H. Sakai for helpful discussion. This work was carried out by the joint research in the Institute for Solid State Physics, the University of Tokyo.
\end{acknowledgments}

\end{document}